# Advancing resistivity–chargeability modeling for complex subsurface characterization using machine learning and deep learning

Adedibu Sunny Akingboye[1, 2, 3, 4], Andy Anderson Bery[1, 2], Hui Tang[4], Ayokunle Olalekan Ige[5],

Obinna Chigoziem Akakuru[6], Gabriel Abraham Bala[1, 2, 7], Mbuotidem David Dick[1, 2, 8]

[1]Geophysics Programme, School of Physics, Universiti Sains Malaysia, 11800 USM, Penang, Malaysia
[2]Earth System Processes and Hazard Modeling Center, Geophysics Programme, School of Physics, Universiti Sains Malaysia, 11800 USM, Penang, Malaysia
[3]Earth Systems, Hazards, & Computational Modeling Center, Department of Earth Sciences, Adekunle Ajasin University, Akungba-Akoko, Ondo State, Nigeria
[4]Helmholtz Centre Potsdam – GFZ German Research Centre for Geosciences, 14473 Telegrafenberg, Potsdam, Brandenburg, Germany
[5]University Canada West, Vancouver, British Columbia V6Z 0E5, Canada
[6]Department of Geology, Federal University of Technology, Owerri, PMB 1526, Imo State, Nigeria
[7]Department of Physics, Federal University Gashua, Nguru Road, Gashua, Yobe State, Nigeria
[8]Akwa Ibom State Polytechnic, 1200 Ikot Osurua, Ikot Ekpene, Akwa Ibom State, Nigeria

**Abstract**

Subsurface lithological heterogeneity presents challenges for traditional geophysical methods, particularly in resolving nonlinear electrical resistivity and induced polarization (IP) relationships. This study introduces a data-driven machine learning and deep learning (ML/DL) framework for predicting 2D IP chargeability models from resistivity, depth, and station distance, reducing reliance on field IP surveys. The framework integrates ensemble regressors with a one-dimensional convolutional neural network (1D CNN) enhanced by global average pooling. Among the tested models, CatBoost achieved the highest prediction accuracy ($R^2$ = 0.942 training, 0.945 testing), closely followed by random forest, while the stacked ML/DL ensemble further improved performance, particularly for complex resistivity–IP behaviors. Overall accuracy ranged from $R^2$ = 0.882 to 0.947 with RMSE < 0.04. Integration with k-means clustering enhanced lithological discrimination, effectively delineating sandy silt, silty sand, and weathered granite influenced by saturation, clay content, and fracturing. This scalable approach provides a rapid solution for subsurface modeling in exploration, geotechnical, and environmental applications.

**Highlights**

1. Novel ML/DL framework enhances resistivity–IP modeling, addressing nonlinear inversion challenges.
2. CatBoost outperforms all models ($R^2 = 0.947$), enabling direct IP estimation from resistivity.
3. 1D CNN with global pooling enhances feature extraction, improving resistivity–IP predictions.
4. KMCA optimally classifies subsurface materials, refining resistivity–IP-based lithologic mapping.
5. A scalable framework supports high-precision subsurface modeling in geologically complex terrains.

**1 Introduction**

Subsurface lithological heterogeneity presents major challenges in geological and geotechnical investigations, often causing ambiguities in geophysical interpretations [1–3]. Variations in soil and rock properties affect fluid flow, mechanical behavior, and resource distribution, making accurate subsurface characterization essential [4–7]. Conventional methods such as seismic refraction, ground-penetrating radar (GPR), and electromagnetic (EM) surveys provide valuable insights but are often constrained by resolution, depth, or sensitivity limitations [8, 9]. Among these, electrical resistivity tomography (ERT) and induced polarization (IP) are widely used for mapping subsurface structures due to their complementary sensitivity to resistivity and chargeability variations [10–12]. However, the resistivity–IP relationship is highly nonlinear across lithological boundaries. Granite, for example, typically exhibits high resistivity but low chargeability, while clay-rich or weathered zones show low resistivity and high chargeability. Saturated or clay-filled fractures further complicate these trends [6, 13].

ERT measures bulk resistivity, aiding in the differentiation of lithological units, assessment of fluid saturation, and detection of structural discontinuities [14, 15]. However, in complex, resistive terrains, ERT alone often fails to resolve key ambiguities [16]. IP complements ERT by measuring the ability of subsurface materials to store and release electrical charge, offering sensitivity to mineralogical composition, clay content, and disseminated sulfides [11]. It is particularly useful for identifying weathered zones, saturated fractures, and conductive clays [17]. Nonetheless, IP surveys are time-consuming, signal-limited, and prone to electromagnetic coupling. Traditional joint ERT–IP inversion techniques further simplify the resistivity–chargeability relationship, often assuming homogeneity across complex geological interfaces [12, 18]. Machine learning (ML) and deep learning (DL) offer data-driven alternatives to overcome these limitations by enhancing resolution, predictive accuracy, and interpretational robustness in heterogeneous settings.

Multiphysics modeling approaches, including the integration of borehole logs and seismic refraction with resistivity–IP data, have reduced ambiguities and improved subsurface resolution [6, 19–21]. Recently, combining ERT–IP modeling with ML/DL techniques has emerged as a promising strategy to address the limitations of deterministic joint inversion.

Methods such as ML-based regressions and k-means clustering analysis (KMCA) enable learning of complex nonlinear relationships in resistivity–IP datasets [13, 18, 22]. However, ML/DL applications in joint resistivity–IP analysis remain underexplored compared to resistivity-only or velocity–resistivity studies [2, 23–25]. Architectures like artificial neural networks (ANN), recurrent neural networks (RNN), convolutional neural networks (CNN), and hybrid ensembles have shown potential in automating geophysical interpretation and reducing uncertainty [2, 26–28]. These AI-driven approaches are particularly valuable in granitic terrains like Penang Island, where complex lithological variability demands real-time, high-resolution imaging.

Penang Island, Malaysia, poses significant geophysical challenges due to its dual granitic plutons, heterogeneous weathering profiles, fractured bedrock, and landslide-prone terrain [29, 30]. In such tropical crystalline settings, resistivity–IP relationships are highly nonlinear, influenced by clay content, variable moisture, and pervasive fracturing [19]. Comparable to the current study, previous studies in the Malaysian granitic terranes, such as Bala et al. [13], applied statistical regression to integrate resistivity and IP datasets from Langkawi, Malaysia, for anomaly detection. However, these methods often underperform in highly resistive terrains where nonlinear interactions dominate. A subsequent study by Bala et al. [31] scaled the approach with ML methods—including simple linear regression (SLR), support vector machines (SVM), k-nearest neighbors (KNN), and CatBoost—using data from Parak, Malaysia. Despite these improvements, the models still lacked the capacity to fully capture complex nonlinearities, particularly those requiring deep learning and advanced ensemble frameworks. To overcome these limitations, this study introduces a novel ML/DL-integrated framework that directly predicts 2D IP models from resistivity data, incorporating depth and station distance as added key inputs. The methodology employs ensemble tree-based regressors, neural networks, and a one-dimensional convolutional neural network (1D CNN) with global average pooling to better resolve nonlinear resistivity–IP interactions. Additionally, unsupervised ML, KMCA is applied to refine lithological discrimination and enhance subsurface classification accuracy.

Beyond methodological innovation, this study has practical implications for Penang Island's evolving landscape. By integrating ML/DL regression, including unsupervised clustering, the framework enables automated classification of lithological units and weathered zones, supporting geohazard mapping, groundwater management, and infrastructure planning. Outputs such as fracture zone delineation, clay layer detection, and weathering front mapping are directly relevant to geotechnical risk assessment and environmental monitoring. Unlike traditional inversion methods, the proposed approach is scalable and adaptive to data-sparse, geologically complex terrains. It offers a rapid, high-resolution alternative for subsurface modeling where field data acquisition is challenging or costly.

## 2. Geological settings

### 2.1 Geological settings and location of the study area

Southeast Asia comprises multiple continental blocks and volcanic arc terranes amalgamated along suture zones, preserving remnants of ancient Tethyan ocean basins [32–34]. The geology of Peninsular Malaysia (Fig. 1a) reflects this complex tectonic history, primarily shaped by the collision between the Western Sibumasu Terrane and the Sukhothai Arc (East Malaya Block) during the Late Triassic [35, 36]. Three principal geological belts—Western, Central, and Eastern—define the region, separated by the Bentong-Raub Suture Zone (BRSZ), a remnant of the Paleo-Tethys Ocean marked by ribbon-bedded cherts, schists, and ophiolitic fragments [32, 37]. These belts represent the accretion of Paleozoic to Mesozoic terranes, with major fault systems such as the Bukit Tinggi, Kuala Lumpur, and Lebir Faults influencing sedimentation, magmatism, and mineralization [37, 38]. The Western Belt, which includes Penang Island, is composed of Ordovician–Permian sedimentary and metamorphic rocks intruded by Late Triassic S-type granites of the Main Range Granite Province [39–41]. The Central Belt, separated from the Western Belt by the BRSZ, contains Paleozoic–Mesozoic volcanic and sedimentary sequences, while the Eastern Belt, part of the Indochina Block, is dominated by Permo-Triassic volcanic rocks and I-type granites [42]. Over 90% of Peninsular Malaysia's plutonic rocks are granitic, with the Main Range and Eastern Granite Provinces marking significant Late Triassic magmatic events driven by crustal melting during tectonic collisions [40, 41]. Structural deformation is evident through thrust faults, folds, and strike-slip systems [35, 38, 43], which control granite emplacement and subsequent geological evolution.

Penang Island (the study area: Fig. 1b), located off Malaysia's northwest coast in the Malacca Strait, is underlain primarily by granitic rocks of the Western Belt, specifically the Main Range Granite Province. The island hosts two major plutonic bodies—the North Penang Pluton (NPP) and South Penang Pluton (SPP)—composed of Late Triassic S-type granites emplaced during the Sibumasu–Indochina collision [39, 40]. These peraluminous, crustal-derived granites share magmatic origins with other tin-bearing granitoids of the Main Range [32, 41, 42]. The NPP includes the Tanjung Bungah, Paya Terubong, and Batu Ferringhi granites, along with the structurally distinct Muka Head microgranite characterized by fine-grained textures and localized deformation [39, 44]. The SPP is dominated by the Batu Maung granite (~80%) and the finer-grained Sungai Ara granite (~20%) [3, 39, 45]. These granites exhibit varying degrees of weathering, influencing soil formation, hydrology, and geotechnical behavior. Quaternary deposits, including the Gula, Beruas, and Simpang Formations, are locally present in lowlands and river valleys but are limited in extent compared to their widespread occurrence on the mainland [46, 47]. These formations consist of fluvial, marine, and alluvial sediments. Penang's heterogeneous granite bedrock, with its variable fracture intensities, overburden thicknesses, and mechanical properties, significantly influences geophysical responses and subsurface processes such as slope stability and groundwater flow [3, 7, 48]. These geological complexities present challenges for conventional geophysical surveys but also provide an opportunity to test and validate advanced AI-assisted resistivity–IP modeling for improved subsurface imaging and predictive accuracy in tropical granitic terrains.

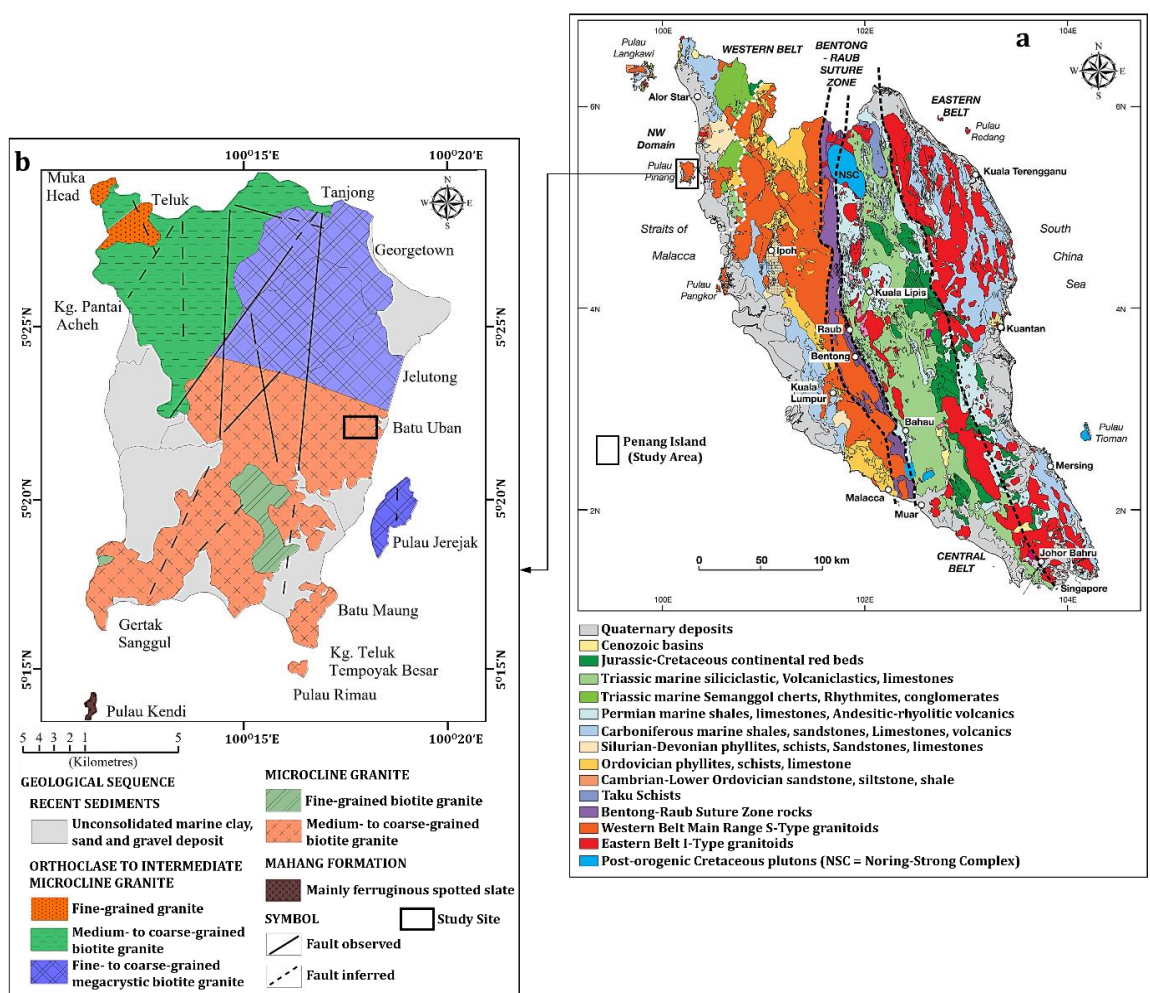

Fig. 1: (a) Detailed geological map of Peninsular Malaysia showing Penang Island (modified after [49]). (b) Geological map of Penang Island showing the study area (modified after [39]).

## 3. Methodology

This study developed a reproducible ML/DL framework for 2D chargeability prediction and lithological classification (Fig. 2). Field data acquisition involved ERT and IP surveys, collocated by station coordinates and depth, with conventional inversion generating baseline resistivity–IP models. Outlier filtering was applied to ensure input quality. Supervised regression models—including linear and polynomial regression, SVM, decision trees, random forest, gradient boosting, CatBoost, artificial neural networks, and a 1D CNN with global average pooling—were trained to predict chargeability from resistivity, depth, and station distance. Model performance was evaluated using $R^2$ score and root mean square error (RMSE), with hyperparameter tuning to minimize error. All implementations used Python (Scikit-Learn and Keras libraries). An unsupervised k-means clustering was then applied to the ML/DL outputs and inversion results for lithological classification, enabling automated mapping of soil–rock interfaces, weathered zones, and fractures. The full dataset and source codes used in this study are openly accessible at https://github.com/ASAkingboye/ML-DL-Resistivity-Chargeability-Modeling.

### 3.1 ERT and IPT field data surveys

To evaluate resistivity and chargeability distributions across surface–subsurface lithological units in granitic terrain, the Universiti Sains Malaysia (USM) Main Campus was selected as the study site. The location provides access to boreholes for lithological constraints and includes diverse surface–subsurface features such as clay/silt layers, concrete slab columns, tree roots, and road sub-base materials, making it ideal for testing resistivity–IP methodologies. Three geophysical survey sites (S1, S2, and S3) were established with traverse lengths of 200 m, 100 m, and 100 m, respectively (Fig. 3; Table 1). This layout was designed to capture detailed near-surface features and broader lateral and vertical variations, despite building obstructions at S3. Field observations identified a silty-to-sandy surficial soil layer overlying granitic bedrock. At S3, the survey line intersected massive tree roots, erosion-induced concrete slabs, and a drainage path, enhancing the detection of subsurface complexity.

To improve model resolution and signal quality, copper electrodes were deployed at S1 and S2, while stainless steel electrodes—offering distinct conductivity and sensitivity characteristics—were used at S3 [17, 50]. Boreholes BH1 and BH2 along S1 (Fig. 3; Table 1) provided lithological constraints, aiding in the calibration of ERT–IP models for soil–rock interface interpretation. Data acquisition was conducted using the Lund Imaging Resistivity System, comprising the ABEM SAS Terrameter 4000 and ES 64–10C Electrode Selector. The Wenner-Schlumberger array was chosen for its high sensitivity to both horizontal and vertical subsurface structures, enabling detailed imaging of lithological variations [14, 51, 52]. Topographic variations along the survey lines were recorded and incorporated into the data processing to ensure accurate subsurface modeling.

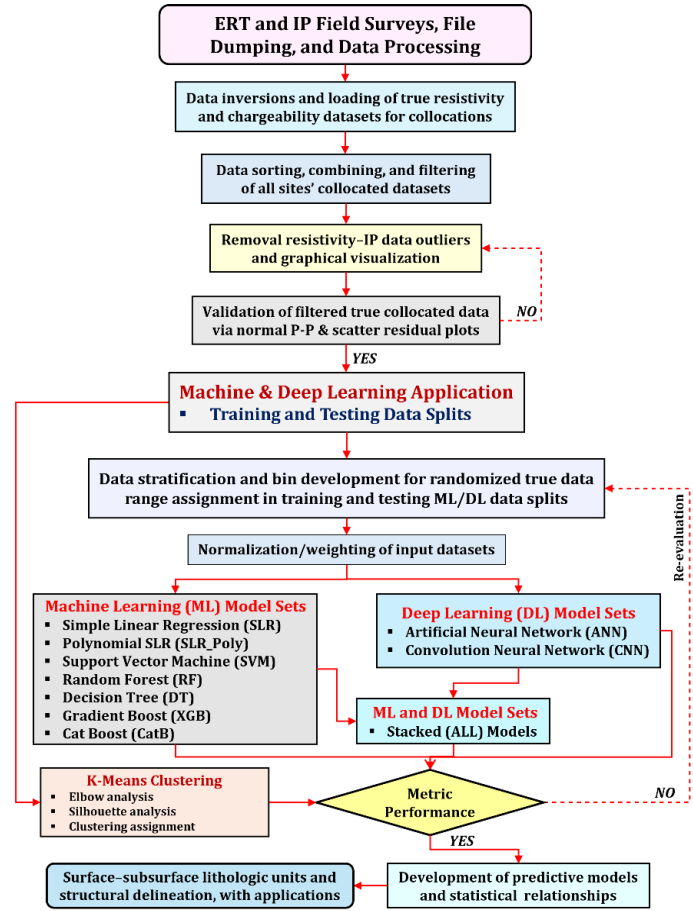

Fig. 2: Research methodological framework illustrating field survey acquisition, data processing, inversion, ERT–IP dataset collocation, filtering, ML/DL training and testing, and KMCA-based modeling approaches.

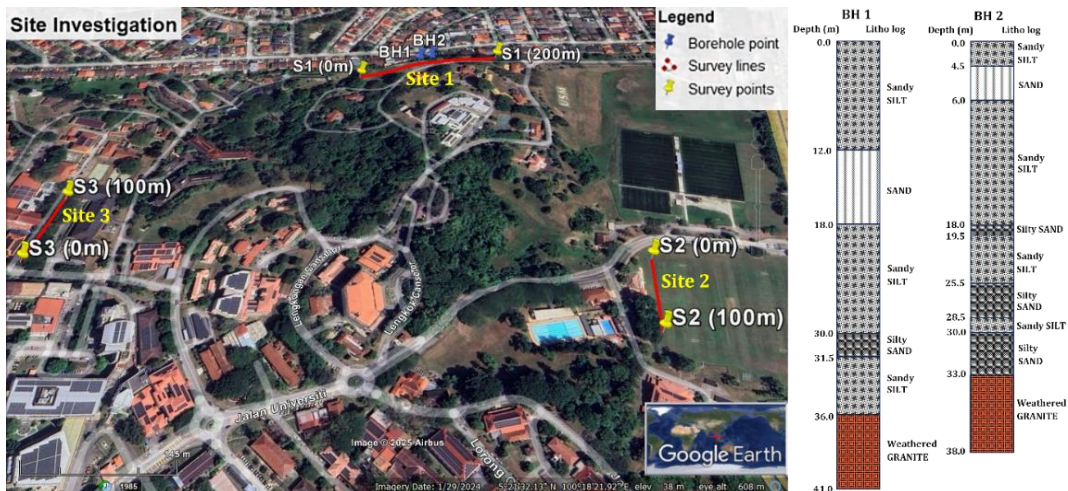

Fig. 3: Aerial geophysical map of the study area within the USM Main Campus, Penang Island, showing the established ERT and IP survey lines and intersected boreholes BH1 and BH2, with their lithological logs displayed on the right side of the map.

Table 1: Summary of survey acquisition site layouts, electrode types, number of boreholes, and inversion software.

| Location | Latitude (N) | Longitude (E) | ERT/IP traverse | Traverse length (m) | Electrode spacing (m) | Approximate traverse orientation | Electrode type | Number of boreholes |
|---|---|---|---|---|---|---|---|---|
| Site 1 | 5°21'43.56" | 100°18'20.09" | 1 | 200 | 5 | W–E | Copper | 2 |
| Site 2 | 5°21'32.58" | 100°18'31.32" | 1 | 100 | 2.5 | N–S | Copper | --- |
| Site 3 | 5°21'32.12" | 100°18'8.13" | 1 | 100 | 2.5 | SW–NE | Stainless-steel | --- |

**Measuring equipment:** Lund Imaging Resistivity Meter—ABEM SAS Terrameter 4000 and ES 64–10C (Electrode Selector), 2 multicore cable reels, copper and stainless-steel electrodes, and GPS
**Array:** Wenner-Schlumberger
**Inversion software:** RES2DINV

### 3.2 ERT and IP data processing, inversion, and modeling

The ERT and IP datasets were processed and inverted using RES2DINV [3, 10] with standard least-squares inversion [51]. A finite-element mesh with four-node elements and L2-norm regularization was applied to better accommodate topographic variations. A damping factor of 0.05 (minimum 0.01) ensured convergence, reducing RMSE to below 10% within five iterations. Initial resistivity and chargeability models served as baselines for minimizing discrepancies between measured and calculated apparent values, improving model accuracy and stability [14, 51]. Maintaining low RMSE was critical to support ML/DL training and ensure robust predictive performance. The final inverted models (Fig. A1a–b) showed high agreement between measured and calculated data, indicating strong signal-to-noise ratios and enhanced subsurface resolution. Investigation depth, typically 25% of the profile length, was controlled by electrode cable configuration and influenced model reliability. Borehole logs from BH1 and BH2 were integrated into the interpretations to refine layer boundaries, weathering profiles, and fracture zones. This integration provided a detailed assessment of subsurface lithological variability, improving the hydrogeological and geotechnical understanding of the site.

### 3.3 ML/DL algorithm development and predictive modeling

The methodological framework (Fig. 2) outlines the workflow for filtering true resistivity–IP datasets, performing co-analyses, and executing ML/DL modeling. The dataset was structured into four columns: X (station distance, m), Y (depth, m), resistivity (ohm·m), and chargeability (msec). Initial visualization identified and removed outliers to preserve model integrity. After validation, multiple ML/DL algorithms were implemented, including Simple Linear Regression (SLR), Third-Order Polynomial Regression (SLR_poly), Support Vector Machine (SVM), Decision Trees (DT), Random Forest (RF), Gradient Boosting (XGB), and Categorical Boosting (CatB), a 3-hidden-layer ANN, and a 1D CNN (Fig. 4). These models were integrated into a stacked ensemble to improve predictive reliability, accuracy, and generalization. Each model was optimized for computational efficiency and robust resistivity–IP correlation analysis. To address data imbalance, a chargeability-based binning strategy was applied to stratify the dataset for balanced training and testing (Fig. 4a). This mitigated bias in underrepresented IP ranges (<30 msec) and ensured robust pairing with the broader resistivity spectrum (up to ~3000 ohm·m). Feature normalization was performed to maintain consistency across varying data scales, enhancing model convergence and performance—particularly for scale-sensitive algorithms—except for DT and RF, which are scale-invariant [2, 53].

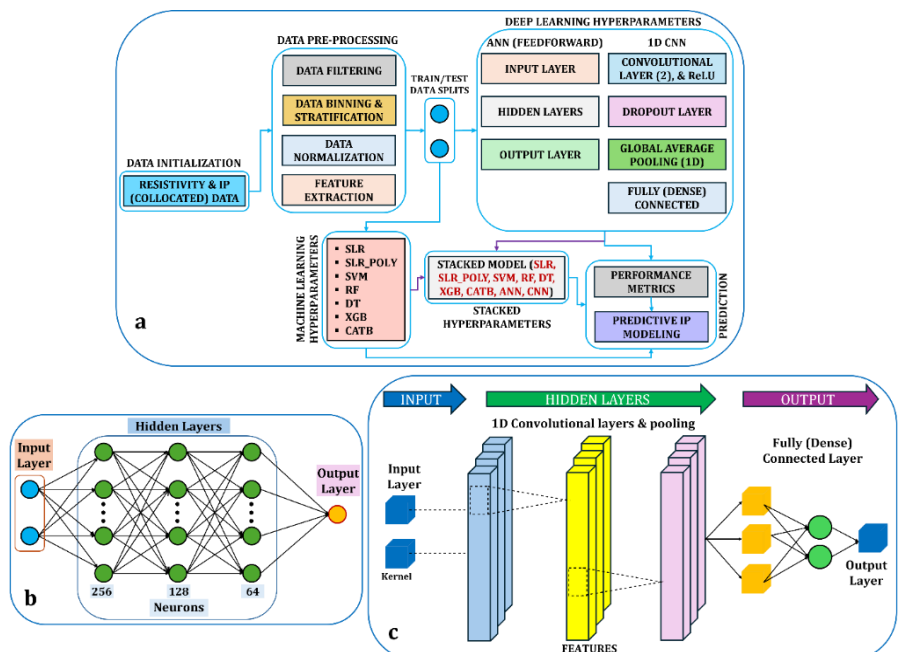

Fig. 4: Proposed algorithm flowcharts for (a) ML/DL-based model, (b) ANN, and (c) 1D CNN architectures.

## 3.4 ML/DL architectures, regressors, and hyperparameters

The ML/DL algorithms were implemented using a structured framework in which 70% (581 samples) of the 830 preprocessed datasets were used for training, and 30% (249 samples) for testing. Table 2 summarizes the model architectures, regressors, layer configurations, and hyperparameters applied for resistivity–IP predictive modeling, treating resistivity as the independent variable and chargeability as the target feature. Iterative parameter tuning during training optimized computational efficiency and accuracy. Model performance was evaluated using standard metrics—MAE, MSE, RMSE, R² score, and adjusted R², capturing both error magnitudes and each model's explanatory strength in mapping resistivity–IP relationships. A summary of the employed models is listed in Table 2.

1. SLR: Serves as a baseline model by establishing a direct linear relationship between resistivity and IP, providing fundamental insights for benchmarking more complex models.
2. SLR_poly: Uses a third-order polynomial to capture nonlinear resistivity–IP trends caused by subsurface complexity, improving predictions where SLR fails [54].
3. SVM: The Support Vector Regression (SVR) with a radial basis function (RBF) kernel used captures subtle nonlinear resistivity–IP variations linked to fractures and weathered zones [55].
4. DT: Uses a single-tree structure with depth and sample constraints for basic resistivity–IP pattern recognition, though best suited as a base learner due to its overfitting tendency [28]
5. RF: Employs an ensemble of 300 decision trees along with other parameters to reduce overfitting and handle complex interactions in resistivity–IP data, especially in geologically mixed or anisotropic zones [56, 57].
6. XGB: Implements a boosting approach with 300 trees and a 0.01 learning rate to iteratively refine predictions, particularly effective in capturing abrupt subsurface property changes. The boosting mechanism allows it to outperform traditional DT in cases where subsurface properties vary significantly over short distances, such as at lithological contacts [58].
7. CatB: An advanced boosting model optimized for both numerical and categorical features, effective in modeling sharp resistivity–IP transitions due to lithological contrasts [59]. The use of L2 regularization, optimized iteration settings, and controlled subsampling further enhances the model's generalization ability, reducing overfitting while maintaining predictive accuracy.
8. ANN: A feedforward multilayer perceptron (MLP) with three hidden layers (256, 128, and 64 neurons) (Fig. 4b), designed to capture complex and nonlinear resistivity–IP relationships. ReLU activations and backpropagation with gradient descent are used for training, with early stopping to prevent overfitting. The ANN effectively models geoelectrical responses linked to mineralization, weathering, and fluid migration [60, 61].
9. 1D CNN: Employs convolutional filters to extract spatial patterns from sequential resistivity–IP data, outperforming ANN in capturing local dependencies [25, 28]. The architecture, Fig. 4c, includes five hidden layers, ReLU activations, a dropout layer rate of 0.2 to prevent overfitting, and a global average pooling layer for dimensionality reduction. A final dense layer with 256 neurons refines features before prediction. This is trained using MSE loss and optimized with Adam, learning nonlinear resistivity–IP patterns with high accuracy [62].
10. Stacked Model: Combines all base models (SLR, SLR_poly, SVM, DT, RF, XGB, CatB, ANN, CNN) using meta-learning to enhance overall prediction accuracy and robustness across variable subsurface conditions.

Table 2: Summary of ML/DL model architectures and key components for the resistivity–IP co-analysis.

| Model | Regressor Type | Layers/Components | Hyperparameters |
|---|---|---|---|
| SLR (Simple Linear Regression) | Linear Regression | Basic regression model | No hyperparameters (uses default scikit-learn settings) |
| SLR_poly (Polynomial Regression) | Polynomial Regression | Uses polynomial features for non-linearity | $Degree = 3$ (Polynomial Features), LinearRegression() as estimator |
| SVM (Support Vector Machine) | Support Vector Regression (SVR) | Uses RBF kernel for non-linear regression | Kernel: RBF, $C = 4.0$, $Epsilon = 0.009$, Gamma='scale' |
| DT (Decision Tree Regressor) | Decision Tree | Single decision tree model | $Max\ Depth = 8$, $Min\ Samples\ Split = 4$, $Min\ Samples\ Leaf = 4$ |
| RF (Random Forest Regressor) | Ensemble (Bagging) | Uses multiple decision trees to reduce variance | $Trees = 300$, $Max\ Depth = 12$, $Min\ Samples\ Split = 4$, $Min\ Samples\ Leaf = 4$, $Features = $ 'sqrt' |
| XGB (Gradient Boosting Regressor) | Ensemble (Boosting) | Uses gradient boosting for improved accuracy | $Trees = 300$, $Learning\ Rate = 0.01$, $Max\ Depth = 6$, $Min\ Samples\ Leaf = 5$ |
| CatB (CatBoost Regressor) | Ensemble (Boosting) | Uses the CatBoost algorithm for better categorical handling | $Iterations = 600$, $Learning\ Rate = 0.05$, $Max\ Depth = 6$, $L2\ Regularization = 5$, $Subsample = 0.8$ |
| ANN (MLP Regressor [feedforward]) | Artificial Neural Network (ANN) | Fully connected neural network based on a feedforward multilayer perceptron | Hidden Layers: (256, 128, 64), Activation Function: ReLU, $Learning\ Rate = 0.0001$, $Max\ Iterations = 5000$, $Early\ Stopping = True$ |
| CNN (1D Convolutional Neural Network) | Convolutional Neural Network (CNN) | Uses Conv1D layers to extract spatial features from tabular data | Conv1D layers: (256, 256), $Dropout = 0.2$, 1D Global Average Pooling, Final Dense: 256, and Optimizer: Adam ($lr = 0.001$), Loss: MSE. Summary: 5 Hidden layers (if $n\_features = 1$), 3 activation functions (ReLU), and a Linear activation function |
| Stacked (Stacking Regressor) | Ensemble (Meta-Learner) | Combines multiple models using a meta-regressor | Base Models: (SLR, SLR_poly, SVM, RF, DT, XGB, CatB, ANN, CNN), Final Estimator: Linear Regression |

## 4. Results and Discussion

### 4.1 ERT and IP inversion models: lithologic units and structural characterization

Figure 5a–d presents the ERT and IP inversion models for Sites S1–S3 and borehole logs (BH1 and BH2), enabling the delineation of subsurface lithologies, interfaces, and weak zones affected by weathering and fracturing. Notably, models from S1 and S2 (Fig. 5a, c), acquired using copper electrodes, show lower RMSE values and distinct anomaly patterns compared to S3 (Fig. 5d), where stainless-steel electrodes and site-specific interferences (e.g., root nodules, concrete, road sub-base) contributed to higher RMSE and variability [17, 50]. Resistivity values span 10–3000 ohm.m, with higher values linked to compacted soil–rock bodies, while chargeability ranges from 0.5–30 msec, indicating clayey to silty materials. Borehole data from BH1 and BH2 at S1 (Fig. 5b) revealed a sequence of sandy silt to silty sand overlying weathered granite bedrock at ~33 m depth, beyond the resistivity–IP imaging limit due to traverse length and electrode spacing. As all sites lie within the USM area and share consistent lithologies, BH1 and BH2 provide adequate validation for the entire study area.

The correlation between borehole logs and resistivity–IP models at S1–S3 (Fig. 5) shows that sandy silt and silty sand exhibit resistivity variations based on moisture content. Saturated layers typically have low resistivity (<200 ohm.m), while dry, compacted zones, especially hardpan clay/silt, exceed 1500–3000 ohm.m. Sandy materials range from ~200 ohm.m (wet) to 3000 ohm.m (dry). Chargeability values aided in refining lithological interpretations, distinguishing water-saturated from clay/silt-rich units. Low chargeability (0.5–6 msec) characterizes sand-rich zones, as seen in S2 (Fig. 5c), reflecting their moisture affinity. Higher chargeability (>6 msec up to 30 msec) marks clay/silt-rich units, especially evident in the central section and station distances 140–175 m of S1 (Fig. 5a) and S3 (station distances 10–20 m, 25–30 m, and 52–90 m; Fig. 5d), where sub-base materials also contribute to elevated values.

Overall, the subsurface lithological units display significant weathering, forming thick residual soil profiles typical of tropical climates [6, 19]. At S3 (Fig. 5d), a highly resistive zone (>1500 ohm.m) between 35–55 m depth with low chargeability (<2 msec) suggests a highly indurated body or a potential buried structure (e.g., a bunker). This interpretation aligns with field observations of vertical water-retention concrete slabs at station distances 33–37 m, 55–57 m, and 70–75 m. Additionally, four massive trees along the profile produced low-resistivity anomalies extending to approximately 5 m depth at station distances 20–35 m, 51–55 m, 70–75 m, and 91–95 m. The presence of these high resistivity and chargeability anomalies at S3 expanded the upper-value range for the ML/DL models. Without these anomalies, model predictions would have been biased toward the lower resistivity and chargeability values observed at S1 and S2 (Fig. 5a, c).

### 4.2 Joint resistivity–IP modeling: filtered true models, visualization, and validation

From modeling observations, depth-referenced ERT and IP models were preferred over topography-based inversion models, yielding ~25% more data points and enabling more efficient large-scale data extraction across heterogeneous terrains. Collocating and filtering resistivity–IP data from combined models (S1, S2, S3; Fig. 5) produced a robust dataset of 830 points for ML/DL training and testing, as detailed in the methodology. The filtered dataset (Fig. 6a) was split into training (70%) and testing (30%) subsets (Fig. 6b) to ensure balanced representation. Key statistics—minimum, quartiles (Q1–Q3), maximum, and standard deviation (STD)—revealed concentration clusters below 700 ohm·m and 5 msec, likely reflecting residual soils, while higher values corresponded to compacted silty sands and weathered granites. The large standard deviations (475.98 ohm·m, 3.33 msec) indicate significant subsurface variability. ML/DL modeling enhanced lithological boundary detection and improved delineation of resistivity–IP ranges, overcoming limitations of conventional inversion in classifying complex subsurface transitions.

A statistical evaluation of the filtered resistivity–IP dataset (Fig. 7a–c) confirmed its suitability for ML/DL predictive modeling. A strong positive correlation (R = 0.93) between resistivity and chargeability indicates shared geological controls, supporting their combined use in subsurface characterization. Regression analysis yielded $R^2$ score = 0.87, standard error = 1.201, Durbin–Watson (D-W) = 1.011, F-statistic = 5539.31, $p < 0.0001$, and collinearity metrics (tolerance and VIF) of 1.000. The large F-statistic confirms substantial variance explanation, while the low Durbin–Watson value indicates independent residuals [63]. The collinearity diagnostics confirm model stability, and the highly significant p-value reinforces predictive reliability [64]. The P–P plot (Fig. 7b) and residual scatterplot (Fig. 7c) further validate normality and homoscedasticity, ensuring unbiased error distribution [65]. To minimize potential bias from upper-range outliers, data transformation and adaptive learning techniques were applied during ML/DL preprocessing, improving model generalization and robustness.

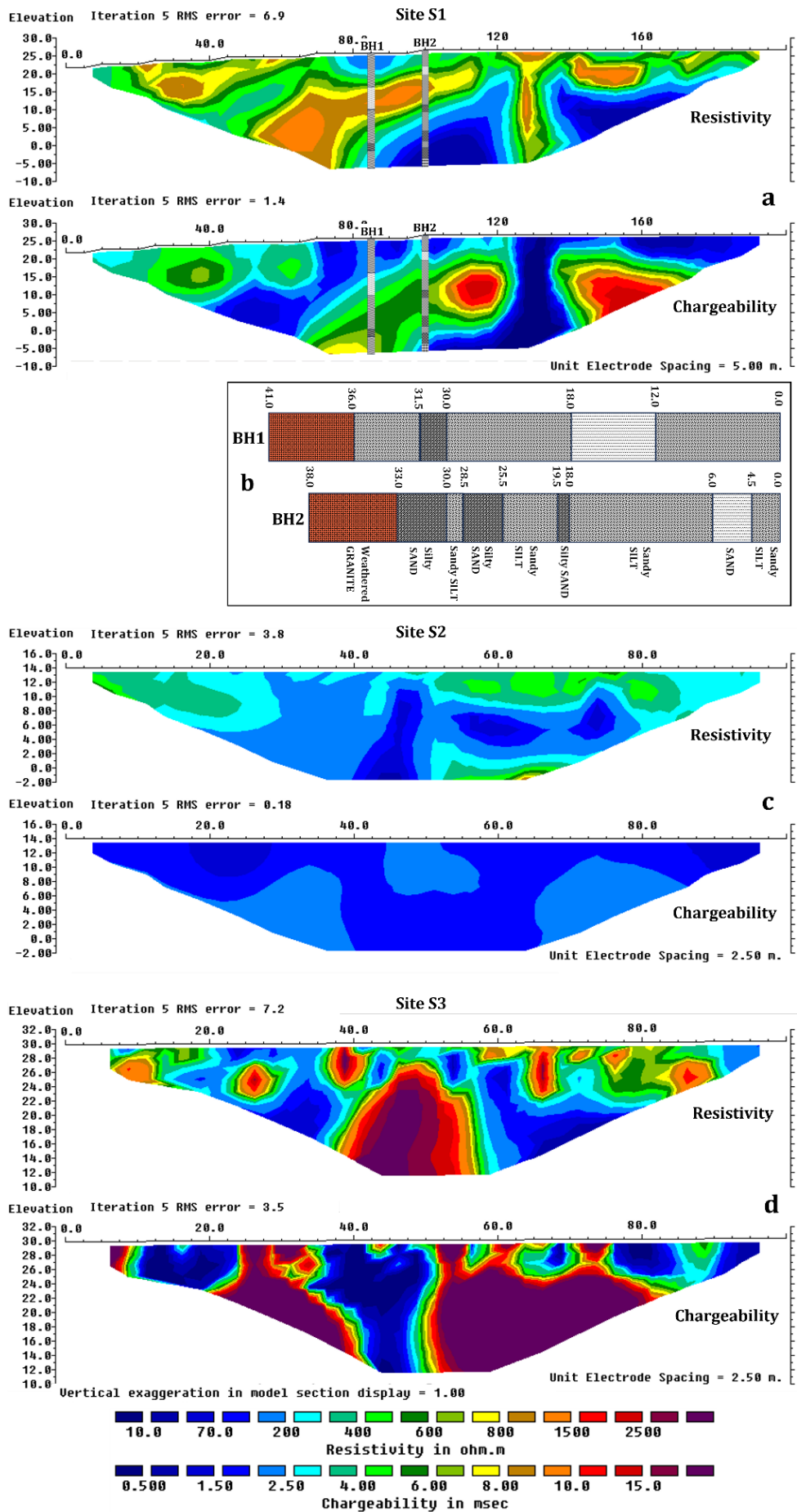

Fig. 5: ERT and IPT inversion models of (a) S1 and (b) borehole lithological logs at stations 85 m and 100 m at S1. ERT and IPT inversion models of (c) S2 and (d) S3. Borehole litholog depths were truncated to match the resistivity and IP model depths.

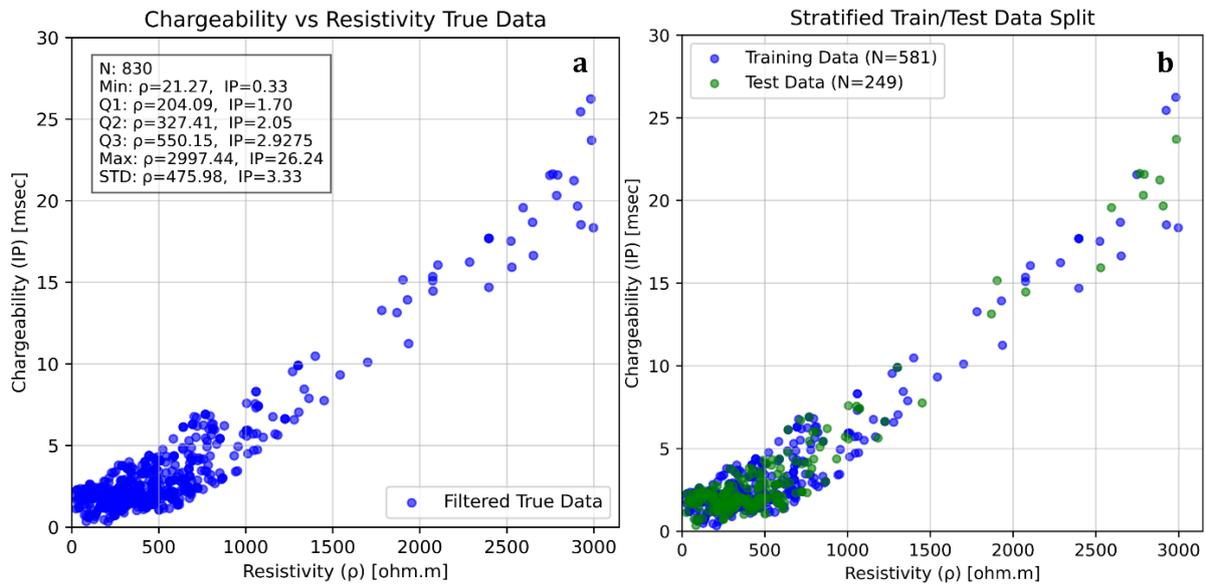

Fig. 6: (a) Filtered true resistivity–IP dataset plot, illustrating the total data points (N=830) along with key statistical parameters, including the minimum, interquartile percentiles (Q1, median [Q2], and Q3), maximum, and standard deviation (STD). (b) Stratified binning of the filtered true resistivity–IP dataset, depicting the categorized data distribution for ML/DL model training (581 data points) and testing (249 data points).

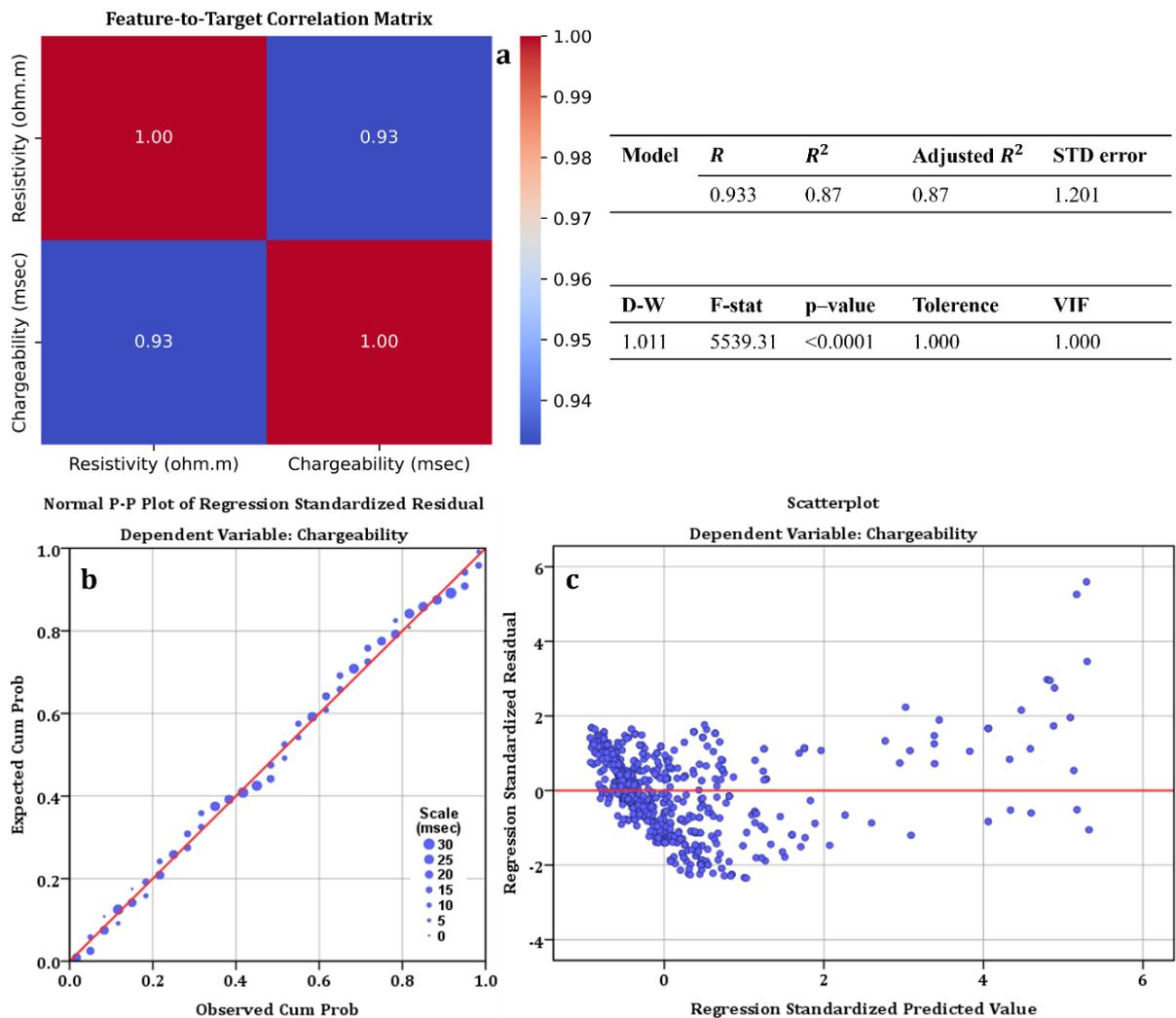

Fig. 7: Accuracy validation plots: (a) feature-to-target correlation matrix with statistical correlation table for resistivity–IP correlation. (b) Normal probability plot (P-P) and (c) scatterplot distribution of regression standardized residual against regression standardized predicted value for the analyzed filtered true resistivity and chargeability datasets.

### 4.3 ML/DL model performance for resistivity–IP prediction

Table 3 summarizes the predictive performance of all ML/DL models, evaluated using MAE, MSE, RMSE, $R^2$, adjusted $R^2$, and $R^2$ difference. SLR recorded the weakest performance, with $R^2$ values of 0.012 (train) and 0.044 (test), far below the statistical benchmark of $R^2$ = 0.87 achieved in Fig. 7. The third-order polynomial regression (SLR_poly) improved significantly over SLR, highlighting the need for nonlinear modeling to capture resistivity–IP complexities. Overall, ML models performed strongly, with $R^2$ values between 0.882 and 0.945 and adjusted $R^2$ between 0.882 and 0.942. SLR_poly and SVM achieved training $R^2$ of 0.909 and 0.908, respectively. Ensemble methods—DT, RF, and XGB—had training $R^2$ of 0.941, while CatB slightly exceeded them at 0.942. Test $R^2$ values were closely matched to training results ($R^2$ differences between −0.035 and 0.002), reflecting effective hyperparameter tuning and stable model generalization. Among DL models, the 1D CNN and ANN achieved training $R^2$ of 0.910 and 0.909, with ANN marginally outperforming in the test set (0.944 vs. 0.943). The stacked model produced the highest test $R^2$ (0.947), despite a lower training $R^2$ of 0.928, indicating superior generalization through ensemble learning [66].

Besides the Stacked (All)—which achieved the highest overall test $R^2$ (0.947), with slightly low train $R^2$ of 0.928 (train); strong generalization supported by low errors (MAE: 0.0241/0.0253, RMSE: 0.0327/0.0330), benefiting from ensemble learning across models—the ML/DL model predictive performance for resistivity–IP relationships (Table 3; Figs. 8a–b, A2a–i) is summarized from best to least performing as follows:

1. CatBoost (CatB): Top standalone model with $R^2$ of 0.942 (train) and 0.945 (test); minimal $R^2$ difference (−0.003); lowest errors (MAE: 0.0205/0.0250; RMSE: 0.0292/0.0337).
2. RF: Matched CatB in test $R^2$ (0.945), with $R^2$ of 0.941 (train); low $R^2$ difference (−0.004); best training error metrics (MAE: 0.0200; RMSE: 0.0294).
3. XGB: $R^2$ of 0.941 (train) and 0.943 (test); small $R^2$ difference (−0.002); consistent errors (MAE: 0.0209/0.0252; RMSE: 0.0294/0.0341).
4. 1D CNN: Best DL generalization with $R^2$ of 0.910 (train) and 0.943 (test); $R^2$ difference (−0.033); low test errors (MAE: 0.0266; RMSE: 0.0342), demonstrating strong spatial feature extraction.
5. ANN: $R^2$ of 0.909 (train) and 0.944 (test); $R^2$ difference (−0.035); test errors (MAE: 0.0268; RMSE: 0.0340), indicating good generalization, though slightly behind CNN in stability.
6. SVM: $R^2$ of 0.908 (train) and 0.943 (test); highest $R^2$ difference (−0.035); moderate errors (MAE: 0.0250; RMSE: 0.0341).
7. DT: $R^2$ of 0.941 (train) and 0.939 (test); $R^2$ score difference (0.002); minor overfitting suggested by higher test RMSE (0.0353).
8. SLR_poly: $R^2$ of 0.909 (train) and 0.942 (test); $R^2$ difference (−0.033); errors (MAE: 0.0273/0.0269; RMSE: 0.0367/0.0344), showing clear improvement over SLR but weaker than DL and ensemble models.
9. SLR: Lowest-ranked model with $R^2$ of 0.882 (train) and 0.914 (test); $R^2$ difference (−0.032); highest errors (MAE: 0.0331/0.0340; RMSE: 0.0418/0.0419), unable to capture nonlinear resistivity–IP relationships.

Table 3: Summarized metric performance for ML/DL and stacked training and test sets for resistivity and chargeability datasets ($N = 830$).

| PISERs/PINNs | Model sets | MAE | MSE | RMSE | $R^2$ | Adjusted $R^2$ | $R^2$ Difference |
|---|---|---|---|---|---|---|---|
| **ML models** | | | | | | | |
| SLR | Train ($N = 581$) | 0.0331 | 0.0017 | 0.0418 | 0.882 | 0.882 | -0.032 |
| | Test ($N = 249$) | 0.0340 | 0.0018 | 0.0419 | 0.914 | | |
| SLR_poly | Train | 0.0273 | 0.0014 | 0.0367 | 0.909 | 0.909 | -0.033 |
| | Test | 0.0269 | 0.0012 | 0.0344 | 0.942 | | |
| SVM | Train | 0.0254 | 0.0014 | 0.0369 | 0.908 | 0.908 | -0.035 |
| | Test | 0.0250 | 0.0012 | 0.0341 | 0.943 | | |
| DT | Train | 0.0204 | 0.0009 | 0.0295 | 0.941 | 0.941 | 0.002 |
| | Test | 0.0261 | 0.0012 | 0.0353 | 0.939 | | |
| RF | Train | 0.0200 | 0.0009 | 0.0294 | 0.941 | 0.941 | -0.004 |
| | Test | 0.0249 | 0.0011 | 0.0337 | 0.945 | | |
| XGB | Train | 0.0209 | 0.0009 | 0.0294 | 0.941 | 0.941 | -0.002 |
| | Test | 0.0252 | 0.0012 | 0.0341 | 0.943 | | |
| CatB | Train | 0.0205 | 0.0009 | 0.0292 | 0.942 | 0.942 | -0.003 |
| | Test | 0.0250 | 0.0011 | 0.0337 | 0.945 | | |
| **DL models** | | | | | | | |
| ANN | Train | 0.0276 | 0.0014 | 0.0367 | 0.909 | 0.909 | -0.035 |
| | Test | 0.0268 | 0.0012 | 0.0340 | 0.944 | | |
| 1D CNN | Train | 0.0269 | 0.0013 | 0.0365 | 0.910 | 0.910 | -0.033 |
| | Test | 0.0266 | 0.0012 | 0.0342 | 0.943 | | |
| **Stacked Model** | | | | | | | |
| Stacked Model (All) | Train | 0.0241 | 0.0011 | 0.0327 | 0.928 | 0.927 | -0.020 |
| | Test | 0.0253 | 0.0011 | 0.0330 | 0.947 | | |

All models, except SLR, achieved RMSE and MAE values well below 0.05, confirming high predictive accuracy [22]. Given that RMSE < 10% is indicative of robust geophysical prediction [22, 59, 67], the results validate the ML/DL framework's effectiveness. The models demonstrate consistent performance across train and test phases, confirming adaptability to complex geological data [2, 68]. Based on test R², RMSE, MAE, and R² difference, the final performance ranking is: CatB ≈ RF > XGB > 1D CNN > ANN > SVM > DT > SLR_poly > SLR. CatB slightly outperforms RF due to its advanced regularization and feature handling capabilities, allowing better generalization than RF's ensemble averaging [58]. Although ANN and 1D CNN underperform during training, both achieve excellent test R² (0.944 and 0.943), highlighting strong generalization. Models like SVM and SLR_poly showed higher test than train R², suggesting favorable behavior in larger test sets. In contrast, ensemble models maintained closely matched R² values, reflecting balanced generalization and stability across datasets.

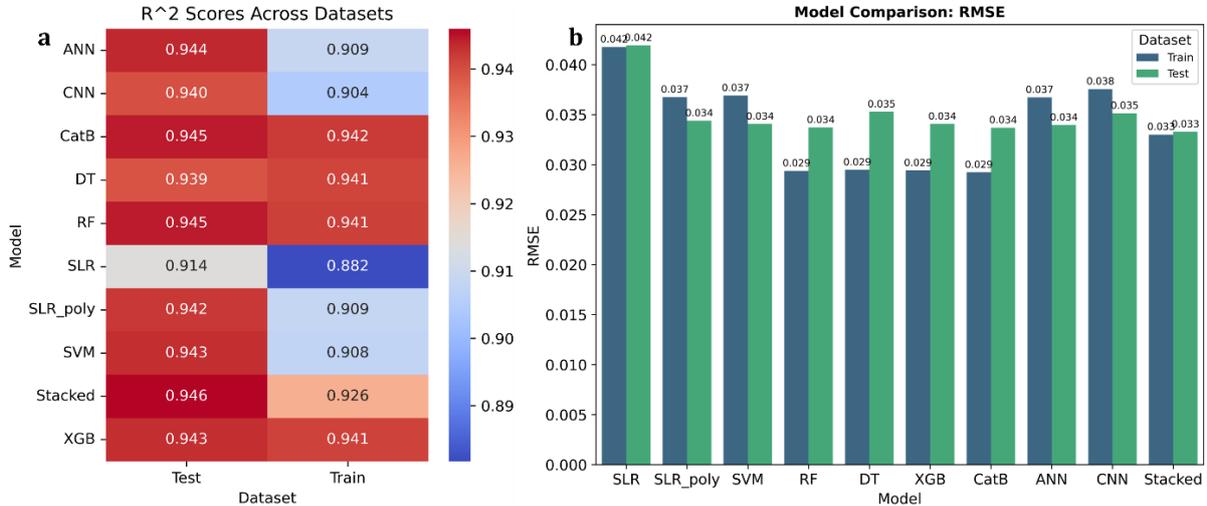

Fig. 8: ML/DL correlation matrix models: (a) $R^2$ and (b) RMSE across the training and test sets.

**4.4 Validation of ML/DL resistivity–IP predictions: model error analysis and comparison**

Residual plots (Figs. A3, 9) show well-distributed errors across models, with no apparent bias or systematic pattern, confirming stable generalization. CatB, RF, and the stacked model produced the tightest residual spreads, reflecting their superior predictive accuracy. ANN and 1D CNN also achieved low residuals, though minor fluctuations in CNN are attributed to localized feature extraction. These findings validate the models' robustness in capturing complex resistivity–IP relationships. The Pearson correlation matrix (Fig. 10) further confirms the inter-model agreement. ANN and 1D CNN exhibit near-perfect correlation (0.999), while CatB strongly correlates with RF (0.993) and XGB (0.996), consistent with their shared ensemble structure. The stacked model achieves the highest overall correlation, demonstrating its capacity to integrate diverse predictive strengths.

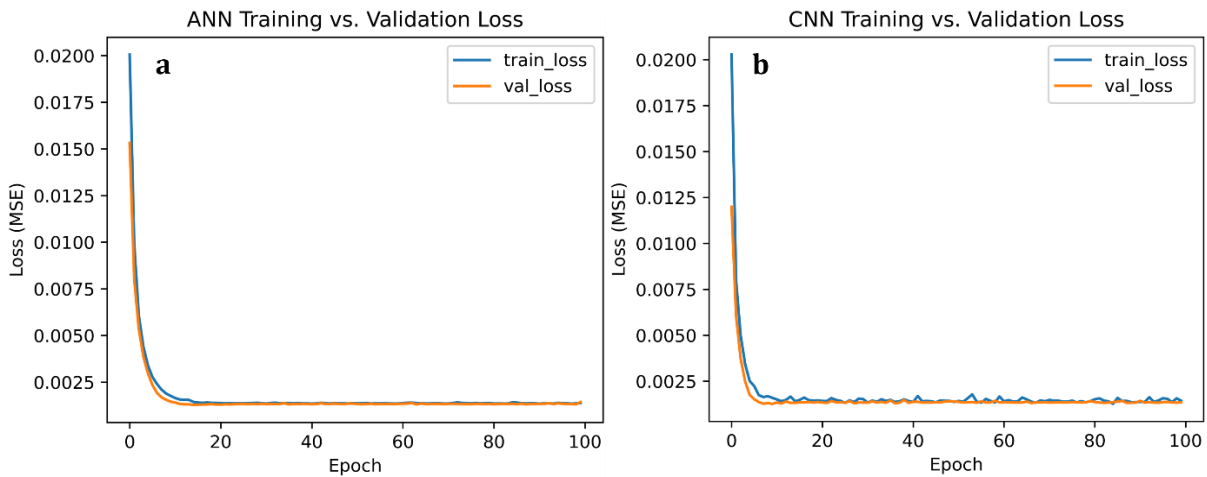

Fig. 9: Loss function curves for (a) 3-layer ANN and (b) 1D CNN model sets.

Normalized predicted vs. actual chargeability values across all models show consistent accuracy (Fig. 11a–b). Aggregated training and test predictions yielded R² values of 0.92 and 0.94, with RMSEs of 0.034 and 0.035, confirming minimal deviation between predictions and actual values. Overall, model performance ranged from 88.2% to 94.7% accuracy (0.882 ≤ R² ≤ 0.947), with RMSE values between 0.0292 and 0.0419 (Figs. 10–11; Table 3). Compared to prior studies in Malaysian terrains (e.g., Bala et al. [13]; SLR R² = 0.8568), this study achieved marked improvements, ranging from 2.52% (SLR) to 9% (stacked model). These results highlight the predictive advantage of ML/DL approaches, particularly when hyperparameters are optimized to balance complexity and accuracy. Except for SLR, all models demonstrated robust performance under complex geological conditions. These error evaluations confirm CatB's superior bias–variance handling, while ANN and 1D CNN showed stronger test-phase generalization, underscoring their strength in modeling nonlinear resistivity–IP relationships.

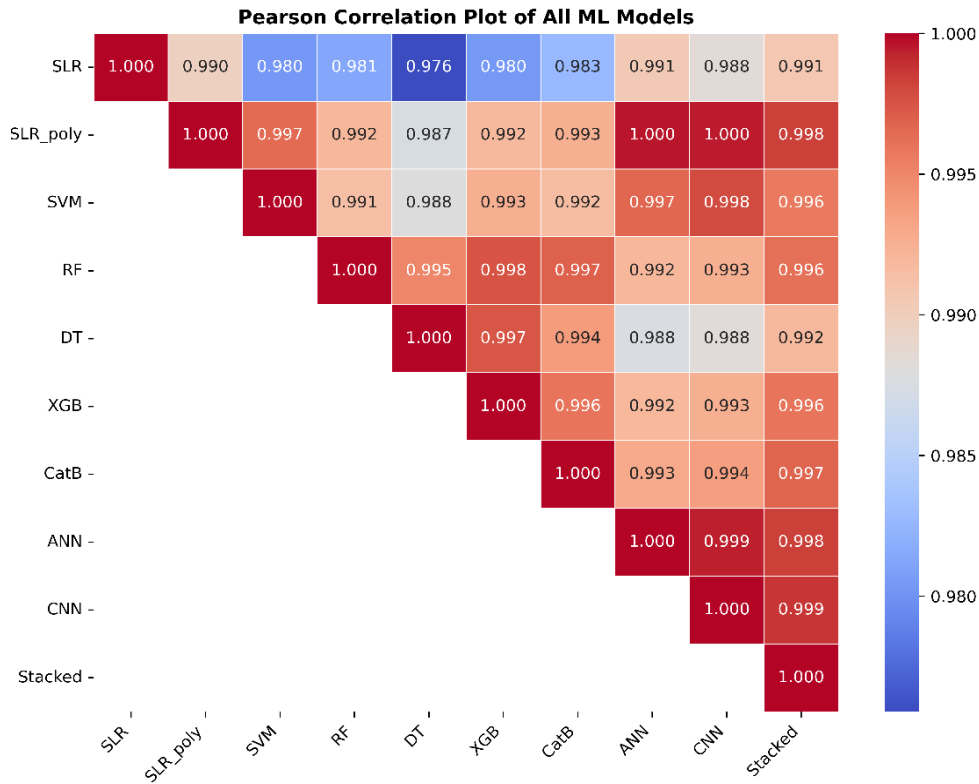

Fig. 10: Pearson correlation matrix model of all the ML and DL models.

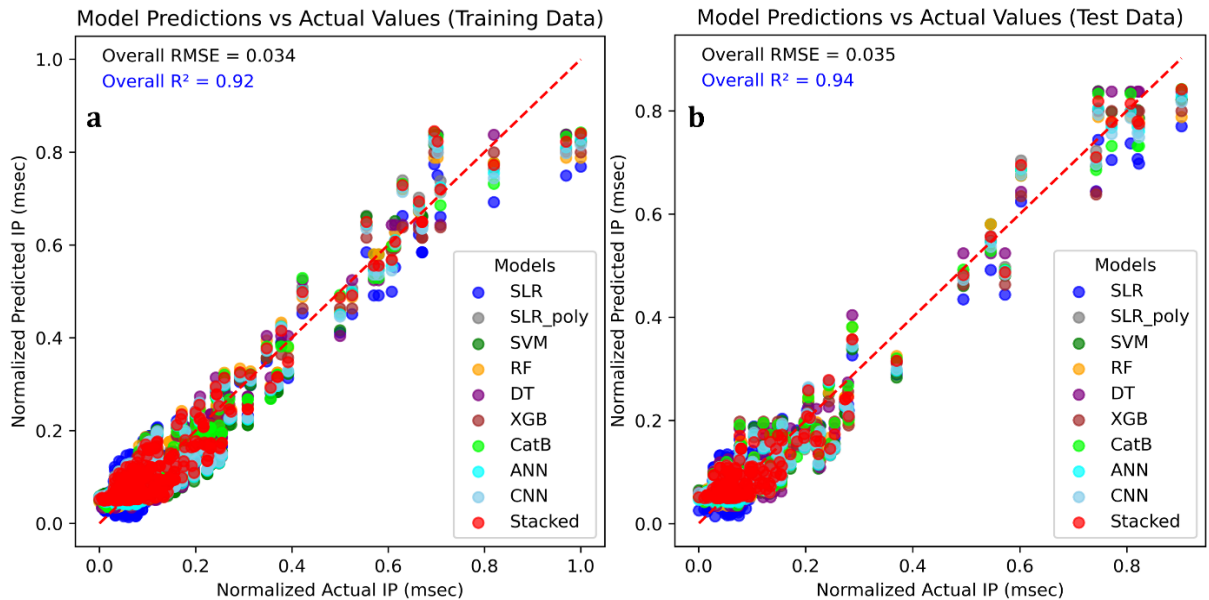

Fig. 11: ML/DL and stacked models of normalized predicted vs actual IP (chargeability) for the overall (a) training and (b) testing datasets.

**4.5 Predictive resistivity–IP modeling for surface–subsurface lithological interpretation**

**4.5.1 ML/DL-based 2D predicted IP models**

To visualize chargeability variations with depth, 2D interpolated ML/DL-based IP models were generated using station distance, depth, resistivity, and chargeability inputs (Fig. 12). These models capture subsurface heterogeneity linked to lithology, moisture content, and clay composition [11, 12, 19]. Electrode layouts at S1 (200 m), S2 (100 m), and S3 (100 m) achieved maximum depths of ~32 m for S1 and <20 m for S2 and S3 (Fig. 5), with outlier removal improving model stability. Across all sites in Fig. 12, low chargeability dominates, reflecting sandy residual soils and weathered granite, while localized moderate-to-high anomalies indicate clay/silt-rich layers or intensely weathered zones with higher charge retention [17, 69]. Some high-IP zones in S1 and S3 lack corresponding high resistivity, partly due to outlier filtering. Tree-based models (DT, RF, XGB, CatB) and DL architectures (ANN, 1D CNN) yielded smoother and more coherent spatial predictions compared to the fragmented outputs of SLR and SLR_poly, while the stacked model further refined spatial patterns by integrating multiple learners. The anomalies are more distinct in S1 due to deeper imaging, whereas S2 and S3, limited to <20 m depth (Fig. 5c–d), lack sufficient resolution to capture deeper features. These results underscore the robustness of ML/DL models in resistivity–IP interpretation, highlighting the importance of algorithm selection and the integration of geological constraints to improve generalization and predictive accuracy in complex granitic terrains [2].

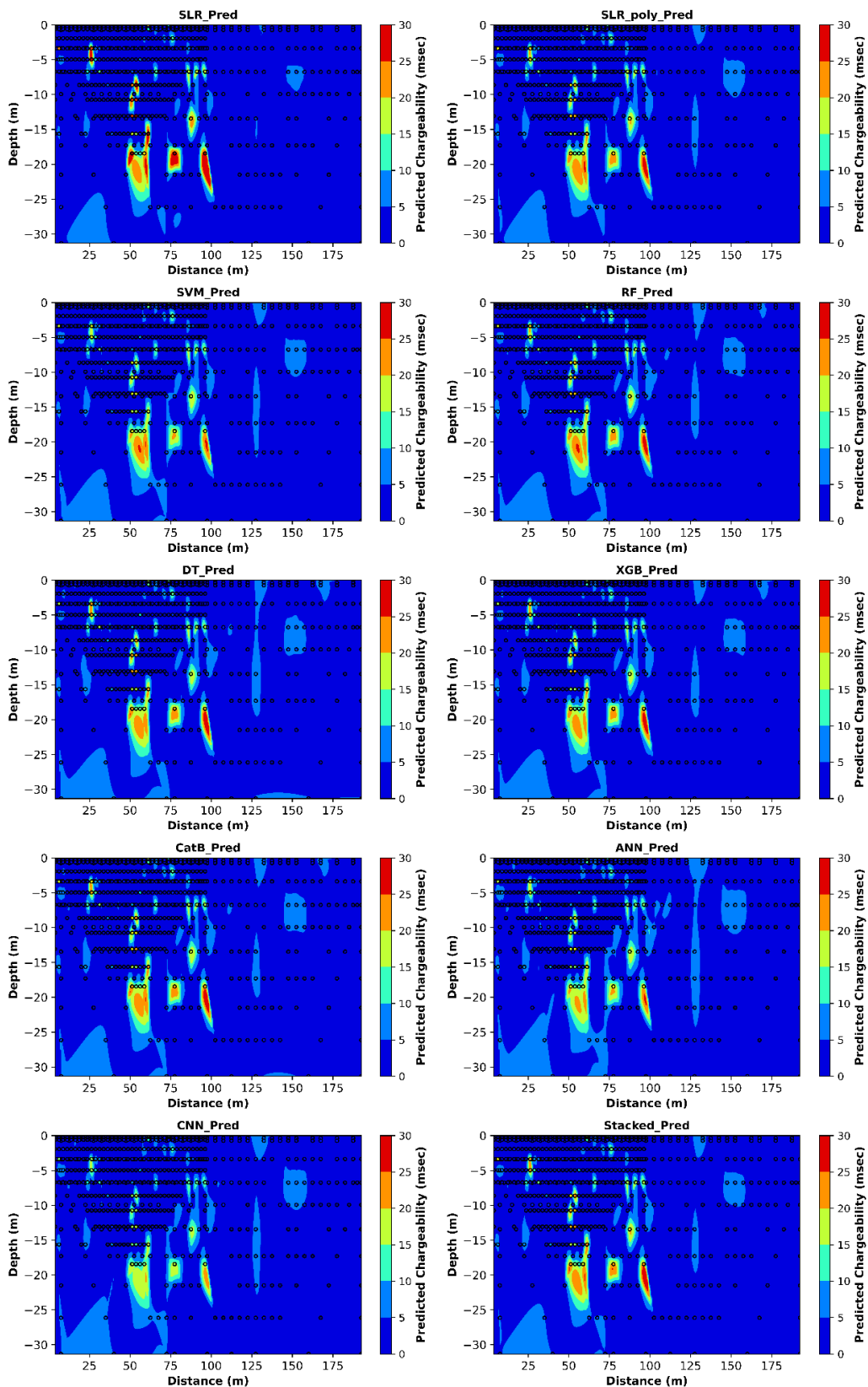

Fig. 12: 2D predicted IP (chargeability) models for all the ML and DL sets.

**4.5.2 ML-based statistical regression predicted IP modeling: 2D exact ERT–IP model visualization**

To develop a practical and high-accuracy statistical regression model for direct IP prediction from resistivity data, the CatBoost (CatB) algorithm—identified as the top-performing ML model—was applied to the filtered true resistivity–IP dataset (830 non-normalized points). The resulting resistivity–IP regression relationship (Fig. 13a) achieved an R² of 0.943, corresponding to 94.3% prediction accuracy. Applying this derived relation to new data yielded predicted IP values, which, when plotted against actual IP values (Fig. 13b), improved the R² to 0.947 (94.7% accuracy), marking a slight 0.004 increase over the initial regression. This trained CatB model was then applied to the inversion-derived IP datasets from S1, S2, and S3 (Fig. 5), producing 2D predicted chargeability models directly from resistivity data (Fig. 14a–c). The comparison between actual and predicted IP models shows strong agreement across all sites, effectively capturing key subsurface features. In S1, the predicted model accurately delineates the central inclined moderate-to-high anomaly and other high-IP zones. In S2, the generally low IP response is well-reproduced, while in S3, the predicted model successfully maps the low IP zone bounded by high chargeability zones, with well-defined boundaries. Minor discrepancies, such as slight shifts in anomaly geometries beneath S1 between 120 m and 140 m, do not affect the overall model integrity. These results validate the robustness of the ML-based statistical regression approach for resistivity–IP modeling, demonstrating its reliability for subsurface chargeability prediction and enhancing the precision of surface–subsurface lithological characterization.

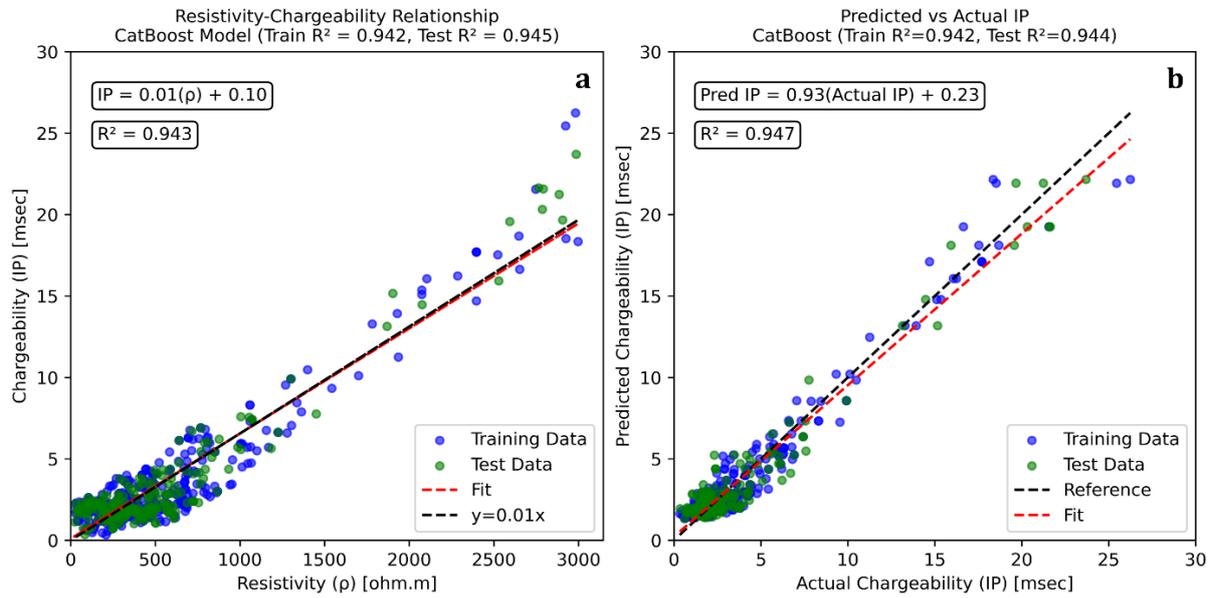

Fig. 13: CatB-developed predictive models: (a) the true resistivity and chargeability training and training sets, and (b) actual vs predicted chargeability (IP) sets.

**4.6 KMCA: insights into lithologic layer classification and subsurface conditions**

The final phase of the methodological framework employed k-means cluster analysis (KMCA), an unsupervised ML technique that classifies data into distinct clusters based on similarities in resistivity and chargeability value [1, 70]. By optimally partitioning the dataset, KMCA effectively differentiated subsurface materials, capturing transitions between soil and rock units [22, 71]. This clustering was independently performed and subsequently integrated with the ML/DL-predicted chargeability models to enhance lithological boundary delineation, improving classification accuracy and subsurface mapping robustness. The KMCA categorizes observations into K clusters by minimizing the objective function ($\varphi$) [22, 72], as defined in Eq. 1.

$$\varphi = \sum_{j=1}^{k} \sum_{i=1}^{N} \left\| x_i^j - \mu_j \right\|^2 \tag{1}$$

where $N$ represents the number of data points, K is the number of clusters, $\mu_j$ is the centroid of cluster $j$, and $x_i$ represents each data point in the datasets.

The algorithm initiates by randomly selecting cluster centers, assigning each point to the nearest centroid, and computing the sum of squared Euclidean distances. The centroids are iteratively updated until $\varphi$ stabilizes, ensuring optimal clustering accuracy. Selecting the appropriate K is critical for meaningful classification, especially when prior knowledge of data structure is limited. To determine the optimal number of clusters, automated validation techniques—the Elbow method and Silhouette method—were employed. The Elbow method evaluates the within-cluster sum of squares (WCSS) to determine the optimal number of clusters [1, 52], with the WCSS expression given in Eq. 2. The Elbow method determines the optimal number of clusters by computing WCSS for various K values and plotting WCSS against K (Fig. 15a). The "elbow point"— where the rate of WCSS reduction notably slows—indicates the optimal K, balancing intra-cluster compactness while avoiding overfitting [70].

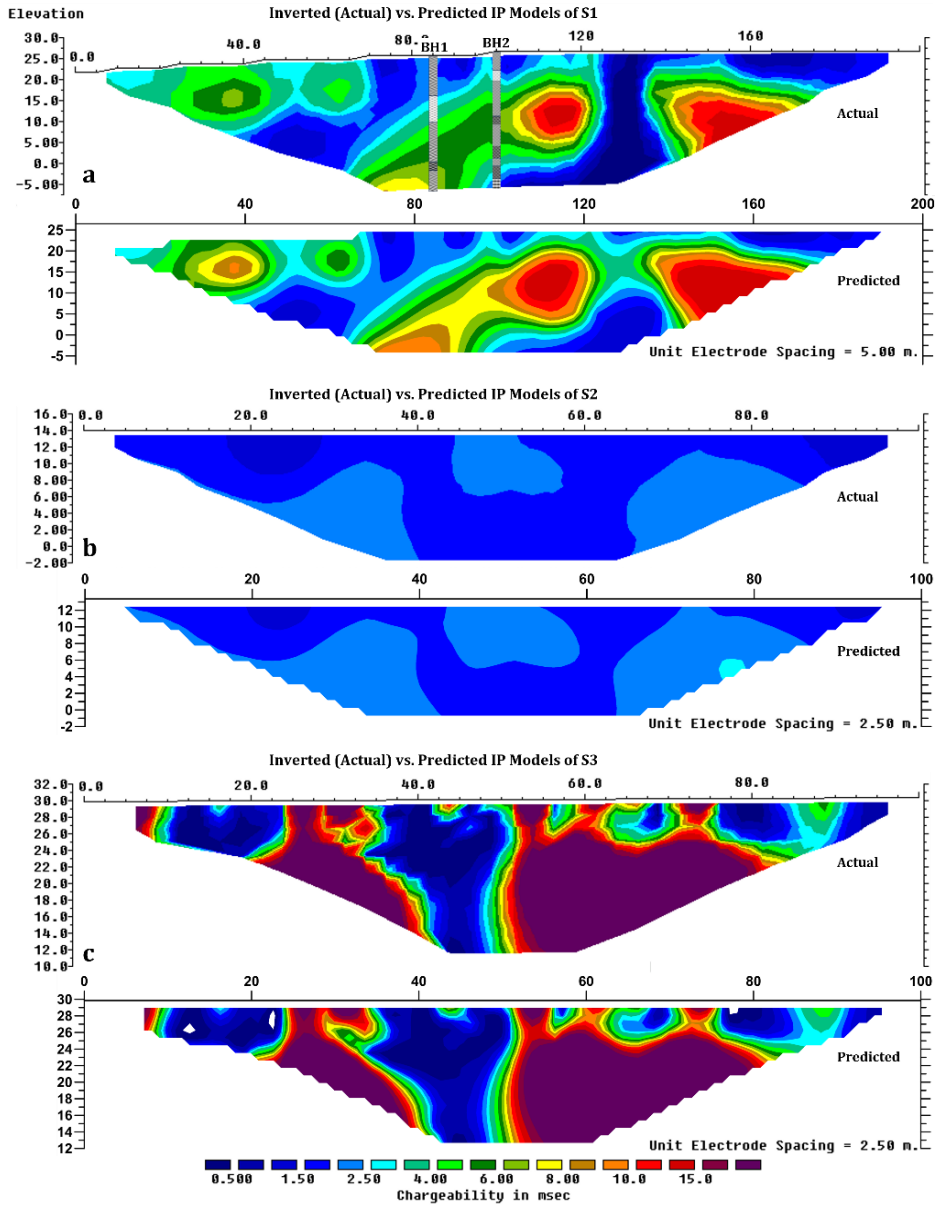

Fig. 14: Actual and predicted IP (chargeability) models of (a) S1, (b) S2, and (c) S3 developed from the inverted (actual) IP and predicted IP relation for the study area.

$$WCSS = \sum_{i=1}^{N}(x_i - \mu_j)^2 \qquad (2)$$

where $a_i$ represents the mean intra-cluster distance, measuring how closely a data point is related to its assigned cluster. Whereas $b_i$ denotes the mean nearest-cluster distance, showing how distinct the data point is from points in neighboring clusters. A lower $a_i$ indicates a strong cluster fit, while a higher $b_i$ indicates well-separated clusters [1, 70].

In contrast, the Silhouette method assesses clustering quality by measuring how well each data point fits within its assigned cluster compared to neighboring clusters [64, 71]. The Silhouette coefficient $S(i)$ for each data point $i$ among $N$ samples is computed using Eq. 3. $S(i)$ ranges from -1 to 1, where values close to 1 indicate well-defined clusters, values near 0 suggest overlapping clusters or unclear boundaries, and negative values signify incorrect clustering or the presence of outliers [1]. The overall Silhouette Index (SI) is obtained by averaging all individual $S(i)$ values, as defined in Eq. 4. By assessing Silhouette scores for different K values, as shown in Fig. 15b, the optimal K is identified, ensuring accurate geophysical classification.

$$S(i) = \frac{b_i - a_i}{max\{a_i b_i\}}, i \in [1, N] \qquad (3)$$

$$SI = \frac{1}{N}\sum_i S(i), i \in [1, N] \qquad (4)$$

Detailed analysis of Fig. 15a–b shows WCSS values decreasing from 10.25 (K=2) to 1.05 (K=10), with corresponding Silhouette scores from 0.853 to 0.391. For meaningful classification, clusters K=3 and K=4 were selected as optimal, yielding average Silhouette scores of 0.624 and 0.511 and WCSS values of 4.79 and 3.20, ensuring balanced cluster compactness and separation [71]. KMCA cluster centroids and their average resistivity–IP values are summarized in Table 4, with clustering results visualized in Fig. 15c, d. These outputs were combined with the ML/DL-predicted 2D IP models and CatB regression models (Figs. 12, 14) to generate a comprehensive lithological classification framework for S1, S2, and S3 (Table 5). This unified classification scheme accounts for variations in resistivity–IP values across similar terrains and enables flexible integration of higher-end values if adopted in future models. The refined KMCA-enhanced framework improves subsurface mapping resolution, offering a robust tool for geotechnical, environmental, and exploration applications in complex tropical granitic terrains.

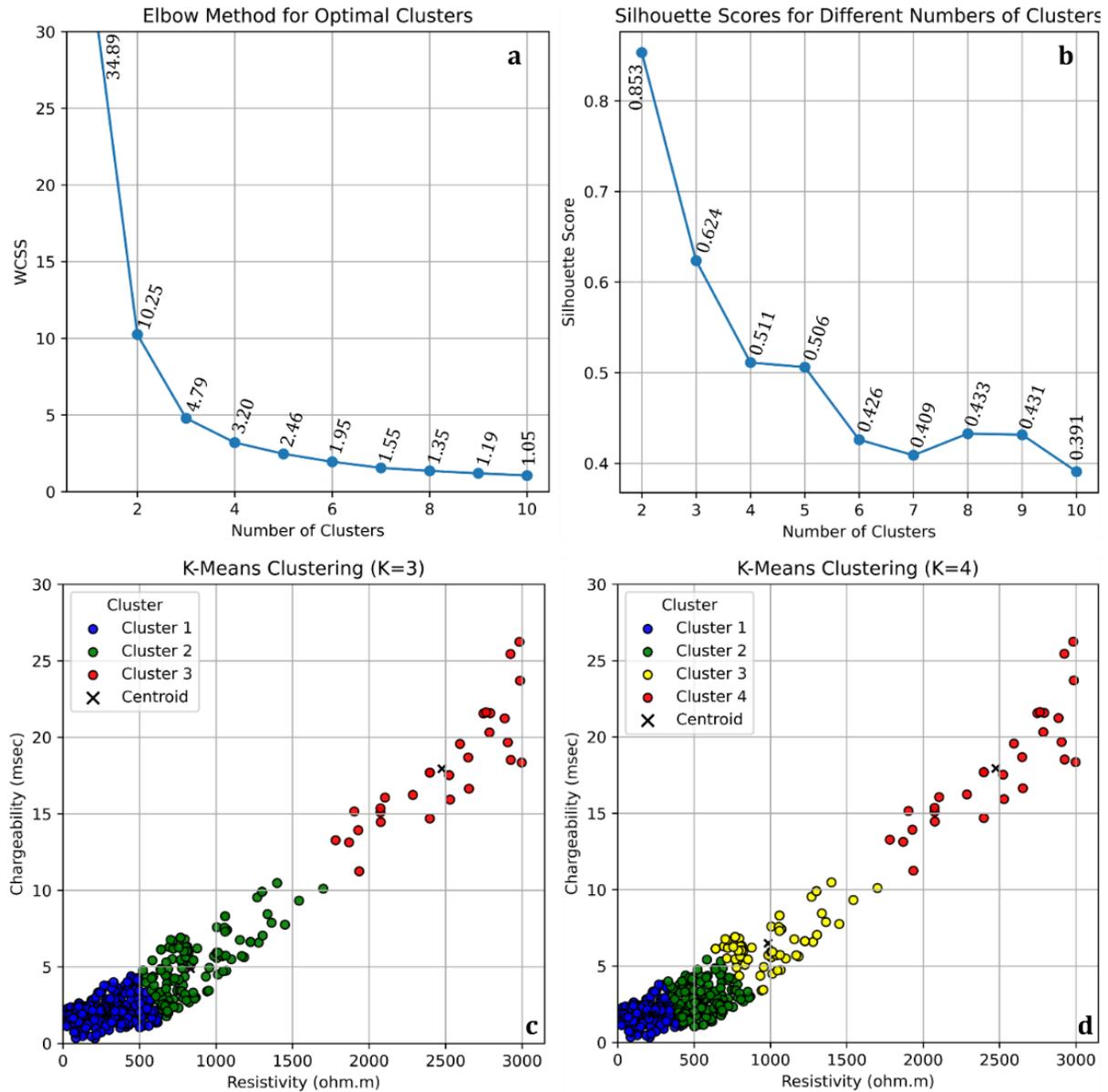

Fig. 15: (a) Elbow method plot for optimal cluster determination and (b) Silhouette scores for different numbers of clusters. KMCA plots: (c) $K = 3$ and (d) $K = 4$ for the study area's lithologic unit delineation.

Table 4: Summarized interpretation of clusters and centroids with the defined specific range of resistivity and chargeability values for $K = 3$ & 4.

| Cluster | Centroids ($K = 3$) | | Centroids ($K = 4$) | |
| --- | --- | --- | --- | --- |
| | Resistivity (ohm.m) | IP (msec) | Resistivity (ohm.m) | IP (msec) |
| 1 | 285.65 | 1.99 | 223.63 | 1.85 |
| 2 | 833.36 | 4.88 | 547.54 | 2.71 |
| 3 | 2475.54 | 17.95 | 984.43 | 6.49 |
| 4 | --- | --- | 2475.54 | 17.95 |

Table 5: Summarized interpretation of the ML resistivity–IP modeling with the lithologic conditions and their implications on the development of sustainable engineering infrastructure and groundwater in the study area.

| Nature of lithologic units | Range of resistivity values (ohm.m) | Range of IP values (msec) | Approximate depth extent (m) | Characteristics and water affinity |
|---|---|---|---|---|
| Topsoil/Residual soil (clay/silt to sand) | <10–700 | 0.5–4.80 | <4.5–12 | Loose, unconsolidated material with variable grain sizes; high porosity and moisture retention, especially in clay/silt-rich zones; moderate to high water saturation potential |
| Highly weathered/fractured granitic unit (silty-to-sandy weathered bodies) | 700–1800 | 2–11 | 4.5–33 | Moderately compact with interconnected fractures; enhanced permeability in silty/sandy zones; moderate-to-high water retention and transmission capacity |
| Relatively weathered/fractured granitic bedrock unit | 1800 to >3000 | 11–30 | >33 | More competent and less porous than the overlying units; retains low to moderate moisture in fractures; limited water storage capacity except in highly fractured sections |

## 5. Conclusions

This study presents a novel, cost-effective, and scalable geophysical modeling framework that integrates ML/DL algorithms to enhance chargeability (IP) prediction directly from resistivity datasets, particularly in geologically complex granitic terrains. By overcoming the limitations of conventional inversion methods, the proposed framework enables rapid, interpretable, and accurate subsurface characterization through the combined analysis of ERT–IP models, ML/DL predictions, chargeability maps, and KMCA-based classifications. The integration of resistivity–IP models with borehole data (S1–S3; BH1 and BH2) reveals predominantly sandy silt to silty sand weathered materials. Variations in resistivity and chargeability are controlled by saturation, clay/silt content, and degrees of weathering or fracturing, confirming the model's ability to resolve complex lithological transitions and subsurface heterogeneities.

The developed ML/DL predictive framework achieved high accuracy ($R^2$ = 0.882–0.947; RMSE = 0.0292–0.0419), effectively delineating subsurface lithologies and moisture–clay distributions. Among the tested algorithms, CatBoost (CatB) and stacked ML/DL ensembles demonstrated superior predictive strength, validating their suitability for complex resistivity–IP modeling tasks. The predicted chargeability models closely matched inverted IP outputs, successfully capturing lithological variations, structural anomalies, and high-chargeability zones with minimal geometric discrepancies and residual errors. KMCA further improved subsurface classification by clustering materials into three to four groups, as verified by the Elbow and Silhouette methods, enhancing geological resolution and supporting geotechnical assessments. The resistivity–IP classification framework developed in this study offers scalability and adaptability across diverse lithologic conditions, accounting for variations in moisture, clay content, and weathered bedrock transitions. While minor challenges remain in modeling abrupt resistivity–IP changes, these can be mitigated using borehole-constrained calibration. Overall, the ML/DL-assisted models provide a reliable and transferable solution for high-precision subsurface investigations, with broad applicability in environmental, geotechnical, exploration, and engineering domains.

**Acknowledgments**

The support of the field technologists at the Geophysics Programme, School of Physics, Universiti Sains Malaysia, who participated in the field data acquisition, is acknowledged. The contributions of all financing bodies that facilitated the successful execution of this research are greatly appreciated.

**Funding**

This research was supported by the Malaysian Ministry of Higher Education (MoHE) under the Fundamental Research Grant Scheme (Grant No.: 203.PFIZIK.6712108), Universiti Sains Malaysia through PKDK VOT29000 and R504-LR-GAL008-0000002198-E140 funding, and financial assistance from Adekunle Ajasin University via TETFund Nigeria. Additionally, the HiDA scientist research grant awarded to the first author for data science at Helmholtz-Centre Potsdam – GFZ German Research Centre for Geosciences, Germany, contributed to this research.

**Data Availability**

All data analyzed during this study are included in this published article. The corresponding authors can make other supporting analyzed data available upon reasonable request. The source codes used in this study are openly accessible at https://github.com/ASAkingboye/ML-DL-Resistivity-Chargeability-Modeling

**Declarations**

**Competing Interest:** The authors declare no known competing financial interests or personal relationships that could have appeared to influence the work reported in this paper.
**Ethical Approval:** All ethical standards have been duly followed during the research.

Consent to Participate: Not Applicable
Consent to Publish: Not Applicable
**Financial interests:** The authors declare no known competing financial interests or personal relationships that could have appeared to influence the work reported in this paper.

**CRediT Author Statement**

**ASA:** Conceptualization, Investigation, Methodology, Software, Coding, Data curation, Resource, Validation, Writing–Draft, Review & Editing, Funding. **AAB:** Data curation, Validation, Resource, Writing–Review & Editing, Funding. **HT:** Data curation, Validation, Resource, Writing–Review & Editing. **AOI:** Data curation, Validation, Resource, Writing–Review & Editing. **OCA:** Writing–Review & Editing. **GAB:** Writing–Review & Editing. **MDD:** Writing–Review & Editing.

**Appendices**

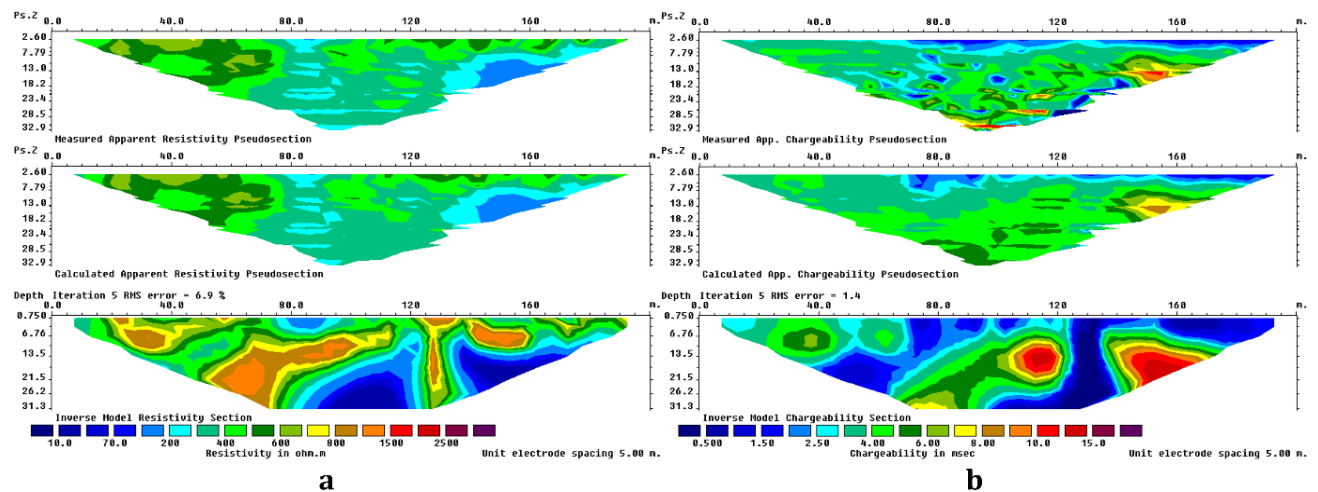

Fig. A1: (a) ERT composite models of measured, calculated, and inverse resistivity sections; (b) IP composite models of measured, calculated, and inverse chargeability sections for Site S1.

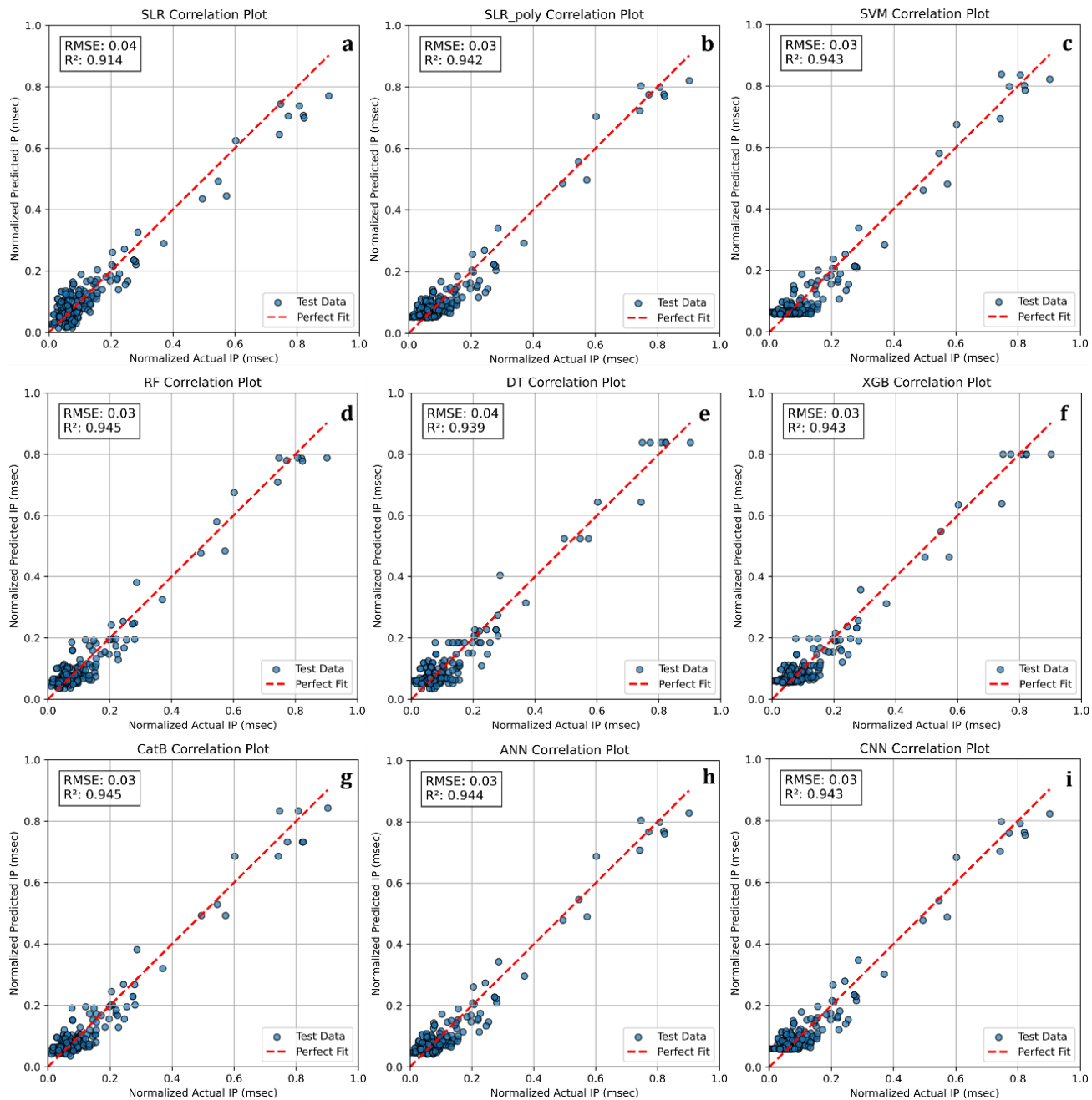

Fig. A2: Correlation plots for ML and DL test predicted chargeability (IP) models of (a) SLR, (b) SLR_Poly, (c) SVM, (d) RF, (e) DT, (f) XGB, (g) CatB, (h) ANN, and (i) 1D CNN sets.

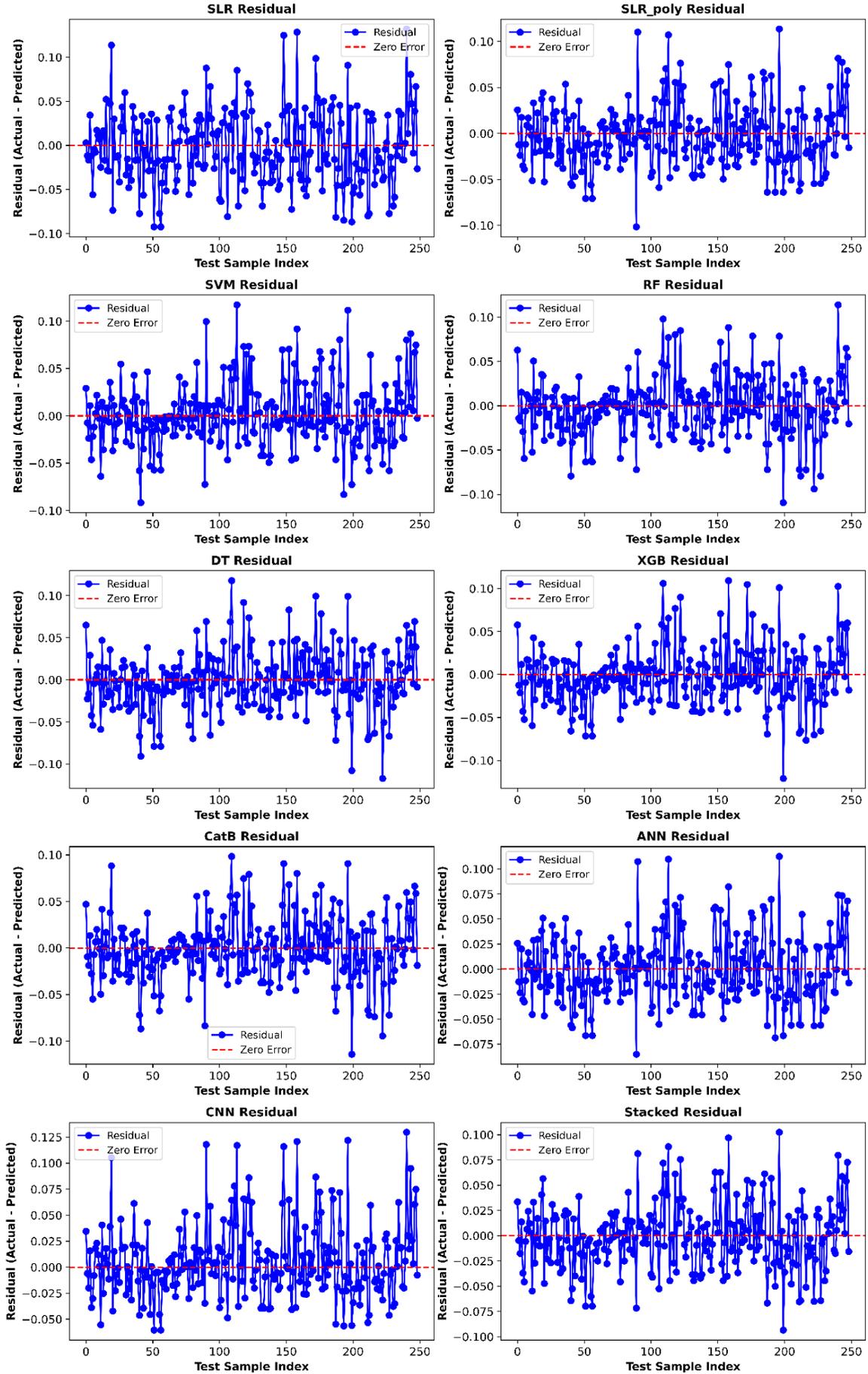

Fig. A3: Residual plots for the ML/DL test prediction models ($N = 249$).